\documentclass[prl,twocolumn,preprintnumbers,footinbib,tightenlines,superscriptaddress]{revtex4}

\pdfoutput=1
\usepackage[english]{babel}
\makeatletter\AtBeginDocument{\let\@elt\relax}\makeatother 
\usepackage{amsmath,amssymb,amsfonts, bm,bbm,slashed, subdepth}
\usepackage{graphicx}
\usepackage{xcolor}
\usepackage{hyperref}
\usepackage{cleveref}
\usepackage{enumerate}
\usepackage{epsfig, subfigure}
\usepackage{setspace}
\usepackage{booktabs, tabularx}
\usepackage{units}
\usepackage{placeins}
\usepackage{multirow}
\usepackage{mathtools}
\usepackage{lipsum,epsfig,dsfont}
\usepackage{soul}
\usepackage{natbib}

\allowdisplaybreaks

\begin{document}
\vspace*{3mm}

\title{Exotic PeVatrons as sources of ultra-high-energy gamma rays }

\author{Andrea Addazi}
\affiliation{Center for Theoretical Physics, College of Physics, Sichuan University,
Chengdu, 610064, PR China.}
\affiliation{INFN, Laboratori Nazionali di Frascati, Via E. Fermi 54, I-00044 Roma,
Italy.}
\author{Salvatore Capozziello}
\affiliation{Dipartimento di Fisica "E. Pancini", Universit\`a di Napoli "Federico
II",  Complesso  Universitario  di Monte S. Angelo, Edificio G, Via Cinthia, I-80126,
Napoli, Italy.}
\affiliation{Scuola Superiore Meridionale, Largo S. Marcellino 10, I-80138, Napoli,
	Italy.}
\affiliation{Istituto Nazionale di Fisica Nucleare, Sezione di Napoli, Napoli,
Italy.}

\author{Qingyu Gan}
\affiliation{Scuola Superiore Meridionale, Largo S. Marcellino 10, I-80138, Napoli,
Italy.}
\affiliation{Istituto Nazionale di Fisica Nucleare, Sezione di Napoli, Napoli,
Italy.}

\begin{abstract}
\noindent
We explore novel classes of exotic astrophysical sources capable of producing ultra-high-energy  gamma rays extending beyond the PeV scale, motivated by quantum gravity scenarios and  dark matter phenomenology. These sources include:
ultra-spinning black hole vortex-string systems; exotic compact objects such as  boson star, axion star and  Q-ball.
Such \textit{Exotica} generate powerful magnetic fields through interactions with millicharged dark matter, enabling particle acceleration mechanisms that surpass the energy limits of conventional astrophysical sources like pulsar wind nebulae  and supernova remnants. We demonstrate that such exotic PeVatrons could be distributed throughout our Galaxy and may be detectable by current (LHAASO, HAWC) and next-generation (CTA) gamma-ray observatories.
\end{abstract}

\maketitle

\section{Introduction}

As it is well known,
the Milky Way (MW) is 
pervaded by relativistic fluxes 
of cosmic rays diffusing through 
 turbulent magnetic fields. 
The cosmic rays spectrum 
extends well beyond the PeV energy 
scale, unequivocably indicating 
the existence of high energy particle 
accelerators in our Galaxy,
i.e. the so dubbed PeVatrons. 

The recent remarkable discoveries made by wide field-of-view gamma-ray detectors, such as those operated by the HAWC, Tibet 
AS$\gamma$ and LHAASO collaborations, have revealed ultra-high-energy (UHE) $\gamma$-rays exhibiting energy spectra beyond the 100 TeV scale \cite{HAWC:2017kbo,ASgamma:2019aml,LHAASO:2021jbf}. This breakthrough has introduced a new perspective into the cosmic electromagnetic spectrum. 
Notably, the identification of more than dozens of Galactic gamma-ray sources surpassing 100 TeV by LHAASO offers substantial observational evidence for investigating the characteristics of PeVatrons. While some of these sources show potential associations with objects cataloged in different astronomical databases, only a limited number are currently linked to known celestial bodies. Moreover, UHE gamma-rays are correlated to hadronic and leptonic channels, opening the possibility 
for a multi-messenger exploration of PeVatrons.

This can suggest the possibility to detect new exotic compact objects (ECOs)
inside the Milky Way, 
inspired by Quantum Gravity (QG) and Dark Matter (DM). In particular, in case of a future detection of ultra high energy gamma rays (UHEGRs) beyond the PeV-scale ($>10$ PeV), there would be no 
way for an explanation from standard astrophysics. 

This could imply the potential discovery of novel exotic compact objects (ECOs) within the Milky Way, drawing inspiration from QG and alternative-to-WIMP DM theories. Specifically, if ultra high-energy gamma rays were detected beyond 10 PeV, conventional astrophysical explanations would likely fall short. For example, energies could be so high that no localized astrophysical source could be related with them so QG curvature corrections could be invoked \cite{GRB_Lamb}. Hence, we introduce the concept of investigating Exotic PeVatrons to address such phenomena.

In particular, there are several hypothetical mechanisms, 
providing for powerful 
cosmic rays accelerators 
from ultra-spinning black holes (BHs)
or BH mimickers. 
Here, we will focus on two main classes  of Exotic PeVatrons (PeVatrons), focusing on UHE $\gamma$-rays: 

\vspace{0.1cm}

i) Exotic Winds (EWinds), 
inspired as a beyond pulsar wind nebulae,
they include several possible new physics systems 
where high energy cosmic ray 
emission is powered by 
 ultra-high spinning BH/ECO with vortex
of DM clouds.

\vspace{0.1cm}

ii) Exotic noave (Enovae),
as a beyond conventional supernovae
or kilonovae, 
are from destabilisation and 
high energy explosion of ECOs 
such as Bosenovae.

Concerning EWinds, 
it is worth to consider the 
case of a ultralight DM boson
cloud surrounding a 
spinning BH or ECO.
The rotating BH/ECO
can cause significant amplification of ultralight bosons through superradiant energy extraction \cite{SR1,SR2,SR3,SR4,SR5,SR6,SR7,SR8}. This process may result in field intensities approaching the Planck scale, providing a 
powerful engine for PeVatronic accelerators in various ways.
Moreover, Vortex String 
can be dynamically formed in BH superradiance 
of dark photon coupled to a
Higgs scalar field \cite{Siemonsen:2022ivj}. 

In case the BH vortex is coupled 
to millicharged particles \cite{Achucarro:1995nu}, 
this can result to the generation 
of a powerful magnetic flux 
for a PeVatronic astroparticle accelerator. 

On the other hand, if 
a spinning BH is reinterpreted 
as Bose-Einstein 
condensate or a gravitational soliton star, 
it can sustain multi-vortex
structures
which,  {\it mutatis mutandis}, 
can also drive 
millicharged dark particles 
around it \cite{Dvali:2021ofp,Dvali:2023qlk}.

Alternatively, in case 
of BH superradiance of vector bosons, 
high energy particles
can be generated 
as parametric resonances 
rather than from accelerating
magnetic fields. 
Which of the mechanisms dominate on 
is decided from the physical boundary conditions and parameters such as
BH mass and spin, DM mass and couplings to ordinary matter, DM local density and composition. 

Recently, it has also been proposed that ultra-light evaporating Primordial Black Holes could generate ultra-high-energy cosmic rays, particularly gamma rays and photons, potentially via the Memory Burden effect, thereby qualifying them as possible PeVatrons \cite{Zantedeschi:2024ram}.

Regarding Enovae, 
they include the so dubbed Bosenovae:
collapse of boson stars (BSs) can lead to explosive emission of a large amount ultra light boson particles \cite{Eby:2016cnq,Levkov:2016rkk,Helfer:2016ljl,Eby:2017xrr,Arakawa:2024lqr, Torres, Cherenkov}, which can convert to cosmic rays. 
Moreover, possible shock mechanism can 
further empower PeVatronic astroparticle acceleration from Bosenovae. 

In the following, we will discuss 
a specific example 
within such a landscape of possible EPeVatrons, selecting a specific 
case of EWinds inspired by
standard pulsar wind nebulae (PWN).

\vspace{0.5cm}

{\it PeV acceleration and UHE Gamma-rays}. 
In pulsar wind nebulae (PWN), electrons accelerated at termination shocks generate ultra-high energy (UHE) gamma-rays through inverse Compton scattering with the 2.7 K cosmic microwave background (CMB) radiation. The conversion efficiency of pulsar rotational energy into relativistic electrons, combined with rapid cooling timescales \cite{WilhelmideOna:2022zmp,Wilhelmi:2024pev}, that enables gamma-ray luminosities approaching the spin-down power
\begin{equation}
L_{\gamma}\sim \dot{E}\, .
\end{equation}
The spin-down luminosity $\dot{E}$ sets the absolute maximum energy for single photons through the relation
\begin{equation}
\label{Egammaa}
E_{\gamma,max}\simeq 0.9\dot{E}_{36}^{0.65}\, \text{PeV}\,.
\end{equation}
This fundamental constraint dominates over synchrotron losses for typical magnetic fields $\simeq 100\, \mu\text{G}$ and $\dot{E}<10^{37}\, \text{erg}\, \text{s}^{-1}$.

The most efficient particle acceleration occurs in uniform electric fields $\mathcal{E}$, where the energy of a nucleus with electric charge $Ze$  gains as
\begin{equation}
E=eZ\mathcal{E}L,
\end{equation}
with  $L$ representing the acceleration path length \cite{Wilhelmi:2024pev}. The induced electric field can be derived from plasma motion through magnetic fields as
\begin{equation}
\mathcal{E}=\frac{\mathbf{U}\times B}{c}\, ,
\end{equation}
where $\mathbf{U}$ is the characteristic plasma velocity \cite{Wilhelmi:2024pev}. Combining these relations yields the {\it Hillas criterion} for maximum energy:
\begin{eqnarray}
E_{max} & = &Z\frac{e}{c}LUB \\ \nonumber
 &\simeq & 0.3 Z\left(\frac{L}{\text{pc}}\right)\left(\frac{U}{10^{3}\,\text{km/s}}\right)\left(\frac{B}{100\,\mu\text{G}}\right)\, \text{PeV}\,.
\end{eqnarray}
 
For PWN systems specifically, the maximum energy depends solely on the rotational energy loss rate
\begin{equation}
\label{Emax}
E_{max}=Ze\sqrt{\frac{\dot{E}}{c}}\,,
\end{equation}
where $\dot{E}$ is the energy loss rate.
The maximal acceleration rate of particles
corresponds to 
\begin{equation}
\label{racc}
r_{acc}^{Max}=\frac{3 \beta^{2} ZeBc}{E}\, ,
\end{equation}
in case of diffusive acceleration for a semi-parallel shock with propagation speed $U=\beta c$ \cite{Amato}.

When energy losses become significant (particularly for PeV electrons), the maximum energy is determined by balancing acceleration and loss rates ($r_{acc}=r_{loss}$) \cite{Wilhelmi:2024pev}. Synchrotron radiation typically dominates the energy losses as
\begin{equation}
r^{sync}_{loss}=\frac{\sigma_{T} c}{6\pi}\frac{B^{2}E}{(m_{e}c^{2})^{2}},
\end{equation}
yielding a loss-limited maximum energy
\begin{equation}
E_{max}^{loss}\simeq 30\, \left(\frac{U}{10^{3}\,\text{km/s}}\right)\sqrt{\frac{100\,\mu\text{G}}{B}}\, \text{TeV}\,.
\end{equation}

This demonstrates that electron acceleration to PeV energies requires the following conditions:
\begin{itemize}
\item Relativistic flow velocities ($\gtrsim 10^{4}$ km/s)
\item Weak magnetic fields ($B<10\,\mu\text{G}$)
\item High spin-down power ($\dot{E}>2\times 10^{35}$ erg s$^{-1}$)
\item Termination shock fields below $3$ mG
\end{itemize}
Such extreme conditions can be naturally achieved by black holes or exotic compact objects (ECOs) with:
\begin{itemize}
\item Masses of $1$-$100\, M_{\odot}$
\item Spin rates comparable to or exceeding pulsars
\item Strong dipole/monopole magnetic fields
\end{itemize}

In  Secs.  II and III, we present concrete black hole-dark matter systems that realize these PeVatron conditions, potentially exceeding conventional PWN acceleration limits. Sec. IV is devoted to discussion and conclusions.

\section{Black Holes, Vortices and Magnetic Flux Structures}

We begin with the Abelian Higgs model in curved spacetime with action given by  \cite{Achucarro:1995nu}
\begin{equation}
S = S_{EH} + \int d^4x\sqrt{-g}\left[\frac{1}{4}F_{\mu\nu}F^{\mu\nu} + D_\mu\varphi^\dagger D^\mu\varphi - V(\varphi^\dagger,\varphi)\right],
\label{Abel}
\end{equation}
where $S_{EH}$ is the Einstein-Hilbert action,  $F_{\mu\nu} = \partial_\mu A_\nu - \partial_\nu A_\mu$ is the $U(1)$ field strength,  $D_\mu = \nabla_\mu + iqeA_\mu$ is the covariant derivative with the  electric charge $qe$ and $\nabla_\mu$ is the covariant derivative with respect to the metric $g_{\mu\nu}$ in curved spacetime. The scalar field can be decomposed as
\begin{equation}
\varphi(x) = v X(x)e^{i\chi(x)}, \quad A_\mu(x) = \frac{1}{qe}[P_\mu(x) - \nabla_\mu\chi(x)]\, . 
\label{PHP}
\end{equation}

This system admits a Nielsen-Olsen vortex solution.  
This can be find solving the Equations of Motions in 
cylindrical coordinates ($\varphi(x) = vX_0(R)e^{i\phi}$) become:
\begin{align}
X_0'' + \frac{X_0'}{R} - \frac{X_0P_0^2}{R^2} - \frac{1}{2}X_0(X_0^2-1) &= 0, \\
P_0'' - \frac{P_0'}{R} - \frac{X_0^2P_0}{\beta} &= 0,
\end{align}
where $\beta = \lambda/2e^2 = m_{scalar}^2/m_{vector}^2$.
While general solutions of these non-linear equations are complicated,
their asymptotic behaviors clearly show that vortex solutions with a non-zero topological winding number
interpolate two different vacua. 
In the limit of $R\rightarrow \infty$,
a Nielsen-Olsen vortex solution is obtained.
Note that $\chi$ given in Eq. \ref{PHP}  is a gauge degree of freedom, which is not dynamical but rather a topological quantity characterizing the winding number of the vortex, i.e. $\oint d \chi=2 \pi n, (n \in \mathbb{Z} )$. Eqs. (12-13) are written in rescaled coordinates with replacement $P_\mu \to n P_\mu$, representing that the
	string has width of order unity and winding number one  (see ref. \cite{Achucarro:1995nu} for more details). 

For black holes with a string/vortex, metric solutions correspond to 
\begin{itemize}
\item Non-rotating case (Aryal-Ford-Vilenkin solution \cite{Aryal:1986sz}):
\begin{equation}
ds^2 \simeq ds_S^2 - r^2(1-4G\mu)^2\sin^2\theta d\phi^2\, , 
\label{SCS}
\end{equation}
 where $\mu$ is the string tensions controlled by the scalar VEV scale. 

\item Rotating case (with small tension $\epsilon \equiv 8\pi G\eta^2$):
\begin{eqnarray}
ds^2 &\simeq& -\frac{\Delta\sigma}{\Gamma}dt^2 
+ (1-2\epsilon\mu)\frac{\sin^2\theta\Gamma}{\Sigma}\left[d\phi + \frac{2aGMr}{\Gamma}dt\right]^2 \nonumber \\ 
&+& \Sigma\left(\frac{dr^2}{\Delta} + d\theta^2\right)
\label{gennn}
\end{eqnarray}
where $\Sigma = r^2 + a^2\cos^2\theta$, $\Delta = \rho^2 - 2GMr$, $\Gamma = \rho^4 - \Delta a^2\sin^2\theta$, and $\rho^2 = r^2 + a^2$\, 
with the dimensionless spin parameter  $a = J/GM^2$.
\end{itemize}

The angular velocity at the horizon is given by
\begin{equation}
\Omega = \frac{a}{2GMr_+} = \frac{a}{2GM(1+\sqrt{1-a^2})}.
\label{OM}
\end{equation}

The vortex configuration leads to quantized magnetic flux through the Aharonov-Bohm effect
\begin{equation}
\label{flu}
\Phi = \frac{1}{qe}\oint P_\mu dx^\mu = \frac{2\pi n}{qe}\, , 
\end{equation}
where $n$ is the winding number and $q e$ is the particle charge. 
Such a flux formula is monopole-like, with magnetic charge $M=1/qe$.

Thus key properties for our PeVatron mechanism are:
\begin{itemize}
\item They exist spinning BH-vortex solutions generating a magnetic flux. 
\item For $q \ll 1$ ($M\gg 1$), large magnetic fluxes are possible. In literature, dark matter candidates with a tiny fractional electric charge  $q \ll 1$ have been proposed, known as millicharged dark matter. Here $q$ is  dimensionless and defined  with respect to an electric unity  charge $e$.
\item Nielsen-Olesen vortices have discrete flux quantization. More general solutions and more complicated models allow continuous flux spectra.
\end{itemize}

The emission power from a BH vortex scales as:
\begin{equation}
P \sim \Phi^2\Omega^2,
\end{equation}
where $\Omega$
is the angular velocity given by 
Eq.\ref{OM}
and $\Phi$ is the magnetic flux in Eq.\ref{flu}.

For an ultra-spinning BH  with mass $M \sim 1-100 M_\odot$, the maximal angular velocity is 
\begin{equation}
\Omega \simeq 1.5 \times 10^5 \left(\frac{R}{1\,\text{km}}\right)^{-1} \text{Hz}\, . 
\end{equation}
This exceeds millisecond pulsar rotation rates ($\sim 10^3$ Hz) by two orders of magnitude, enabling powerful PeVatron emission when coupled to millicharged dark matter in the galactic halo.
On the other hand, the main amplifier for the magnetic flux is the inverse millicharge dependence of $\Phi$.  

In Fig.1, we show how an emission power of $P\sim 10^{37}$erg/s
can be reached for several BH parameters (mass, spin, angular velocity and the charge of millicharge particles).
This demonstrates as a BH-vortex can provide a viable mechanism for PeV and beyond photons. Let us stress again that the mass spectrum of BH considered  is not far from $10\, M_{\odot}$, we do not need for supermassive BHs or for micro PBHs. 
Such a model is minimal which basically introduces only a milli-charged light dark scalar field beyond the SM. 

In the next section, we discuss alternative PeVatrons to vortex-BHs.

\begin{figure*}[htbp]
	\centering
	\includegraphics[width=0.48\linewidth]{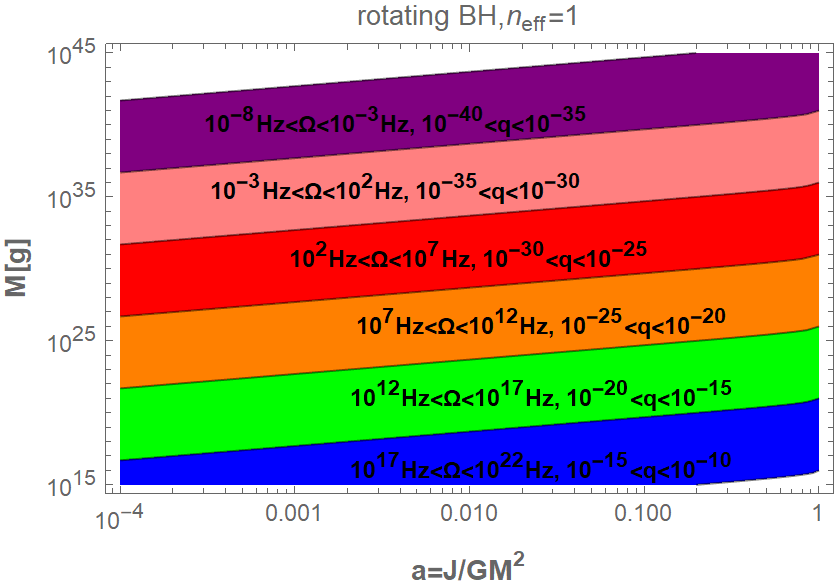}
	\caption{ 
		Emission power $P = 10^{37}$ erg/s in BH parameter space. The parameter $\Omega$ denotes the angular velocity and $q$ the charge of the millicharge particle. Different color represents different combination of $\Omega$ and $q$.  }
	\label{PeVBHFig}
\end{figure*}

\section{Solitons as ECOs}
In this section, we examine soliton-like exotic compact objects (ECOs), including BSs,  axion stars and Q-balls with millicharged dark matter. When these objects rotate, they form vortices that trap quantized magnetic flux. During the spin-down process, they emit SM electric particles in the surrounding plasma or millicharged dark matter particles, creating a novel mechanism for particle acceleration that can produce PeVatron-scale emissions.

\subsection{Boson Stars}

A BS represents a theoretical astrophysical compact object that forms through Bose-Einstein condensation of bosonic particles under self-gravitating conditions (see Ref.\cite{Visinelli:2021uve} for a review on this subject). The equilibrium configuration of such a star arises from a delicate balance between competing forces: gravitational collapse  counteracted by quantum pressure effects (including the Heisenberg uncertainty principle) and, in many models, by repulsive self-interactions within the bosonic field. We analyze a charged complex scalar field $\phi$ with self-interactions in a curved spacetime background, described by the action of Eq.\ref{Abel}. 
 In a renormalizable QFT, the only allowed interaction term corresponds to a four-point vertex, resulting in an effective potential
\begin{equation}
    V(|\phi|)=\frac{1}{2}m^2|\phi|^2-\frac{1}{4}\lambda |\phi|^4,
\end{equation}
where $m$ denotes the particle mass and $\lambda$ represents the self-interaction coupling constant. The non-interacting limit corresponds to $\lambda=0$, with $\lambda>0$ ($\lambda<0$) giving rise to repulsive (attractive) self-interactions. Following standard nomenclature, we refer to the non-interacting case as a ``mini boson star'' (mini-BS) and the interacting case as a ``massive boson star'' (massive-BS). The field strength tensor $F_{\mu\nu} = \partial_\mu A_\nu - \partial_\nu A_\mu$ describes the $U(1)$ gauge field $A_\mu$, with the covariant derivative given by $D_\mu = \nabla_\mu - i(qe) A_\mu$. In our framework, we consider particles with millicharges significantly smaller than the elementary charge $e$, parameterizing the charge as $q e $ where $q \ll 1$.

In the non-relativistic regime, the self-gravitating Bose-Einstein condensate is governed by the Gross-Pitaevskii-Poisson (GPP) system. This framework finds broad application in modeling superfluid phenomena in astrophysical contexts, particularly in neutron star cores. The GPP system admits stable soliton solutions due to its $U(1)$ symmetry, which ensures conservation of particle number $N$. 

For rotating configurations and $\lambda>0$ (Higgsed-phase), the solution takes the form of a Nielsen-Olesen type vortex characterized by an integer winding number $n = 1,2,3...$, leading to quantized angular momentum:
\begin{equation}
    J = nN.
\end{equation}
Treating the rotating soliton as a rigid body with mass $M$ and radius $R$, we obtain the angular velocity
\begin{equation}
    \Omega = \frac{J}{I} \simeq \frac{nN}{MR^2} \simeq \frac{n}{mR^2},
\end{equation}
where $I \simeq MR^2$ represents the moment of inertia and we have used the relation $N \simeq M/m$ in the final equality. 

Remarkably, the vortex with winding number $n$ can trap a quantized magnetic flux 
\begin{equation}
    \Phi = \frac{n}{qe} \, .
\end{equation}
This implies that substantial magnetic fluxes can be sustained when the particle charge $q e$ is sufficiently small -- a scenario naturally realized with millicharged dark matter. During the spin-down phase of the rotating BS, this configuration can eject electric particles with considerable energy. Following \cite{Contopoulos:2005rs}, the emission power scales as:
\begin{equation}
    P \simeq \Phi^2\Omega^2 \simeq \frac{n^4}{q^2 e^2 m^2 R^4}.
    \label{eqP}
\end{equation}
This powerful emission mechanism can potentially produce standard model particles (such as electrons) through interactions in the vortex atmosphere surrounding the exotic compact object.

To estimate the emission energy, we must first determine the geometric properties of rotating BSs, particularly their mass-radius relation. Building upon the comprehensive analysis of non-rotating neutral BSs presented in \cite{Chavanis:2011zi}, we propose the following mass-radius relation for rotating solutions:
\begin{equation}
    R = \frac{(1+n)^\alpha}{GMm^2}\left(1 + \sqrt{1+\frac{GM^2\lambda}{(1+n)^\beta}}\right),
    \label{eq-mass-radius-BS}
\end{equation}
where $R$ and $M$ denote the star's radius and mass, respectively.  Inspired by the mass-radius relation of rotating Axion star given in ref. \cite{Davidson:2016uok}, the parameters $\alpha$ and $\beta$ are introduced by hand in  Eq. \ref{eq-mass-radius-BS}  to account for the rotating states of the BSs. When $n=0$,  Eq. \ref{eq-mass-radius-BS} reduces to the non-rotating case of the BSs' mass-radius relation obtained in    ref. \cite{Chavanis:2011zi}. The appropriate values of $\alpha$ and $\beta$ will be determined in large n limit, as discussed later.  In addition, we take $\lambda \geqslant 0$.  While ref. \cite{Chavanis:2011zi} identifies an additional branch for attractive interactions ($\lambda<0$), we exclude this unstable solution from our rotating case analysis.

Our formulation relies on several well-justified assumptions for this order-of-magnitude analysis:

\begin{itemize}
    \item The millicharged nature of the constituent particles as well as the Yukawa ranges considered allow us to neglect electrodynamical potentials. 
    \item Geometric deformations due to rotation remain $\mathcal{O}(1)$, as evidenced by the extremal Kerr black hole case where the outer horizon radius is half its Schwarzschild counterpart. 
    \item Equation \eqref{eq-mass-radius-BS} remains valid even in the strong gravity regime. 
\end{itemize}

In the strong gravity limit ($R \approx GM$), we obtain the critical mass $M_{\text{crit}}$:
\begin{itemize}
    \item For mini BSs ($\lambda=0$): $M_{\text{crit}} = 1/(Gm)$, consistent with the Kaup limit $0.6/(Gm)$.
    \item For massive BSs: $M_{\text{crit}} = \sqrt{\lambda}/(\sqrt{G}m)^3$, matching the numerical result $0.1\sqrt{\lambda}/(\sqrt{G}m)^3$.
\end{itemize}
As remarked above, the objects of our interest correspond to the non-zero positive $\lambda$ parameter. 

The winding number dependence follows from \cite{Davidson:2016uok}, where $\alpha=3/2$ and $\beta=3$ for large $n$. Numerical studies \cite{Ontanon:2021hbg} confirm the proportionality $M_{\text{crit}} \propto n$, supporting another choice of $\alpha=\beta=2$ for the linear approximation. This simplification agrees with earlier treatments \cite{Lee:1986ts,Ferrell:1989kz}. Since we perform qualitative analysis of millicharged dark matter BSs, thus we simply take  $\alpha=\beta=2$ in the numerical calculation.

\begin{figure*}[htbp]
		\centering
\includegraphics[scale=0.7]{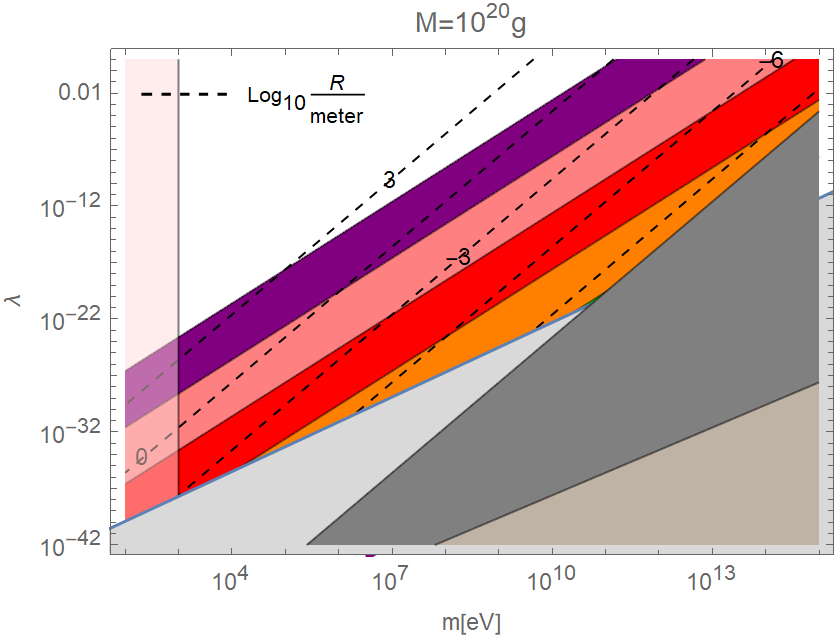}
\includegraphics[scale=0.7]{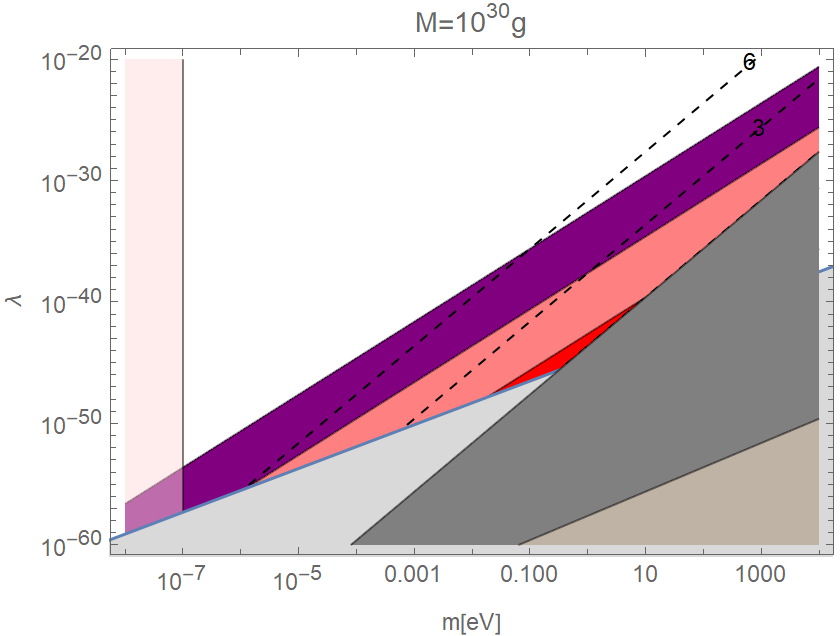}
\includegraphics[scale=0.7]{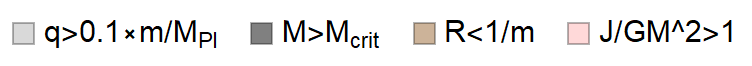}
\caption{ 
Contour plot of the spin-down emission power $P = 10^{37}\,\mathrm{erg/s}$ 
        for rotating BSs with repulsive self-interactions ($\lambda > 0$). 
        The vortex winding number is fixed at $n = 1$. The colored regions (purple, pink, red, orange, green, blue) 
        correspond to the same parameter ranges shown in Fig.~\ref{PeVBHFig}. }
\label{FigBSRep}
\end{figure*}

By combining the preceding expressions, we can express the emission power as either $P(n,m,\lambda,q, M)$ or $P(n,m,\lambda,q, R)$, enabling systematic exploration of the parameter space. 

Figures \ref{FigBSRep} displays contour plots of $P = 10^{37}\,\mathrm{erg/s}$ in the $\lambda$-$m$ plane for repulsive self-interactions, respectively. The plots feature several characteristic regions:
\begin{itemize}
    \item \textbf{Dark gray region}: Parameter space where $GM \gtrsim R$, corresponding to black hole solutions.
    \item \textbf{Light gray region}: Parameter space $q > o(0.1)m/M_{pl}$, where the electrodynamical contribution can not be neglected and hence Eq. \ref{eq-mass-radius-BS} fails.
    \item \textbf{Brown region}: Excluded regime where $R \lesssim 1/m$, violating the minimal size requirement for Bose-Einstein condensate formation.
    \item \textbf{Semi-transparent red region}: Super-spinning configurations with $a = J/(GM^2) > 1$ that exceed the Kerr bound.
\end{itemize}

The super-spinning region warrants special consideration, as such rapidly rotating ultra-compact objects would exhibit strong frame-dragging effects in their ergoregions, potentially leading to instabilities \cite{Cardoso:2007az}. While a detailed stability analysis lies beyond our present scope, we indicate this parameter space through translucent shading for completeness.

For BSs with repulsive self-interactions (Fig.~\ref{FigBSRep}), the parameter space enabling PeVatron emission exhibits several key features:
\begin{itemize}
    \item The viable region expands for smaller BS masses ($M$) and larger bosonic particle masses ($m$).
        \item For $M = 10^{20}\,\mathrm{g}$, PeVatron emission occurs with  possible charge maximum to $q \sim 10^{-20}$.  Whereas for $M = 10^{30}\,\mathrm{g}$, achieving PeVatron energies requires extremely small charges: $q \sim 10^{-40}$--$10^{-30}\,$.
\end{itemize}



\subsection{Stueckelberg Axion Stars}

An alternative possibility for generating a mass to abelian gauge bosons is through 
a Stueckelberg rather than Higgs mechanism
\cite{Stueckelberg:1938zz,Stueckelberg:1938hvi,Ruegg:2003ps}. The Stueckelberg mechanism provides a gauge-invariant description of massive gauge fields by introducing a compensating real (pseudo)scalar field $b$. For a $U(1)$ gauge field $A_\mu$, the key steps are as follows:
\begin{itemize}
    \item Start with a massless gauge field $A_\mu$ transforming as:
    \[
    A_\mu \rightarrow A_\mu + \partial_\mu \lambda.
    \]   
    \item Introduce a Stueckelberg scalar $b$ that shifts under gauge transformations:
    \[
    b \rightarrow b - m \lambda.
    \]    
    \item Construct the gauge-invariant combination:
    \[
    X_\mu = A_\mu + \frac{1}{m} \partial_\mu b.
    \]    
    \item The Proca mass term becomes gauge-invariant:
    \[
    \mathcal{L}_{\text{mass}} = \frac{m^2}{2} X_\mu X^\mu.
    \]
    In unitary gauge ($b = 0$), this reduces to $\frac{m^2}{2} A_\mu A^\mu$.
\end{itemize}

In some theories (e.g., string compactifications or Green-Schwarz string anomaly cancellations \cite{Svrcek:2006yi}), axions can play the role of Stueckelberg fields.
Thus we will consider here the case of a axions for a Stueckelberg mechanism providing mass to the photon (or to the dark photon), with the characteristic periodic potential
\begin{equation}
    V(a) = m^2f^2\left(1 - \cos\left(\frac{b}{f}\right)\right),
    \label{eq-Cos}
\end{equation}
where $f$ represents the decay constant. The non-perturbative
effects reduce the continuous gauge shift transformation 
to a discrete one 
$b\rightarrow b+2\pi f$. 
In the non-relativistic regime, the real scalar field can be decomposed into a complex field, permitting non-topological soliton solutions with conserved particle number. 

For the specific case where $mf \simeq (100\,\mathrm{MeV})^2$, this corresponds to the QCD axion originally proposed to solve the strong CP problem. 
We emphasize that Stueckelberg Axion differs from the Peccei-Quinn one related to a global symmetry rather than a gauge one.
However, since $q \ll 1$, this millicharged axion maintains nearly identical mass-coupling relations to its neutral counterpart.

The potential in Eq.~\eqref{eq-Cos} admits a Taylor expansion for $b/f < 1$ as
\begin{equation}
    V(b) = \frac{1}{2}m^2 b^2 - \frac{1}{4!}\frac{m^2}{f^2}b^4 + \frac{1}{6!}\frac{m^2}{f^4}b^6 + \mathcal{O}(b^8),
    \label{eq-V-AS}
\end{equation}
where we explicitly show the series truncation error. For $b/f \ll 1$, truncation at the quartic term suffices, yielding the {\it dilute axion star} solution. 
This corresponds to a BS with repulsive  self-interactions $\lambda \simeq m^2/f^2$, represented in Fig.~\ref{FigBSRep}.

For realizing a PeVatron we also introduce 
an extra dark millicharged particle $\phi'$ coupled to 
the $X_{\mu}$ as 
\begin{equation}
\label{Xmuj}
\mathcal{L}_{int}=(q e) X_{\mu}J_{\phi'}^{\mu}\, , 
\end{equation}
and a mass $m_{phi'}\simeq m$.

The gauged axion star can sustain a neutral quantized vortex configuration, 
similar to superfluids, where the axion field \( b \) plays the role of the 
vortex phase.
In the vortex core, 
$b=0$, 
while asymptotically 
\[
b = 2\pi n f,
\]
and the gauge field \( A_{\theta} \sim n/r \) screens the \( \partial_{\theta}a/m \) term
 ($\theta$ is the angle variable in cylindrical coordinates), making the energy finite.

Assuming that the axion star is surrounded by a dark cloud of millicharged particles \( \phi' \), the neutral vortex can transfer angular momentum to the \( \phi' \) cloud, creating a charged vortex that sustains a magnetic flux. The resulting mechanism is analogous to the Nielsen-Olesen vortex solution in the Abelian Higgs model.

\begin{figure*}[htbp]
\includegraphics[scale=0.7]{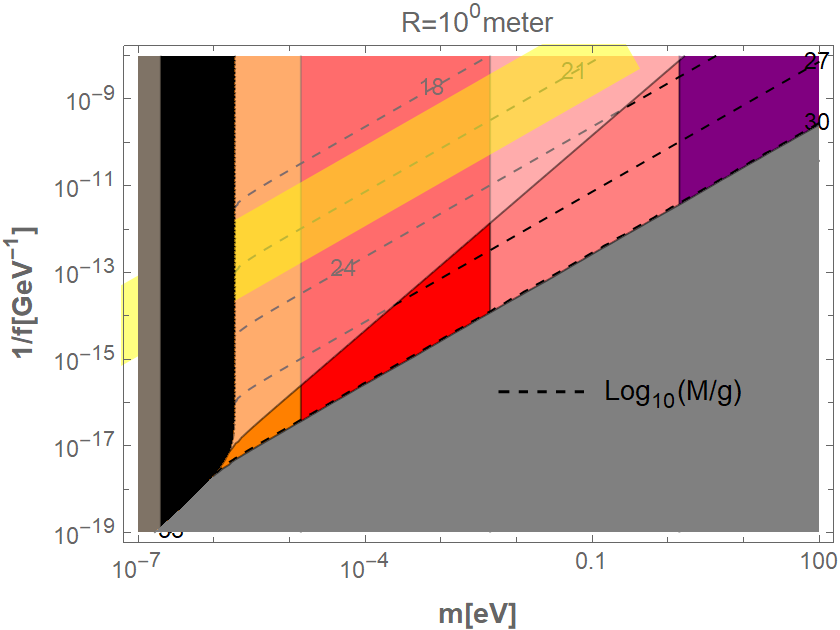}
\includegraphics[scale=0.7]{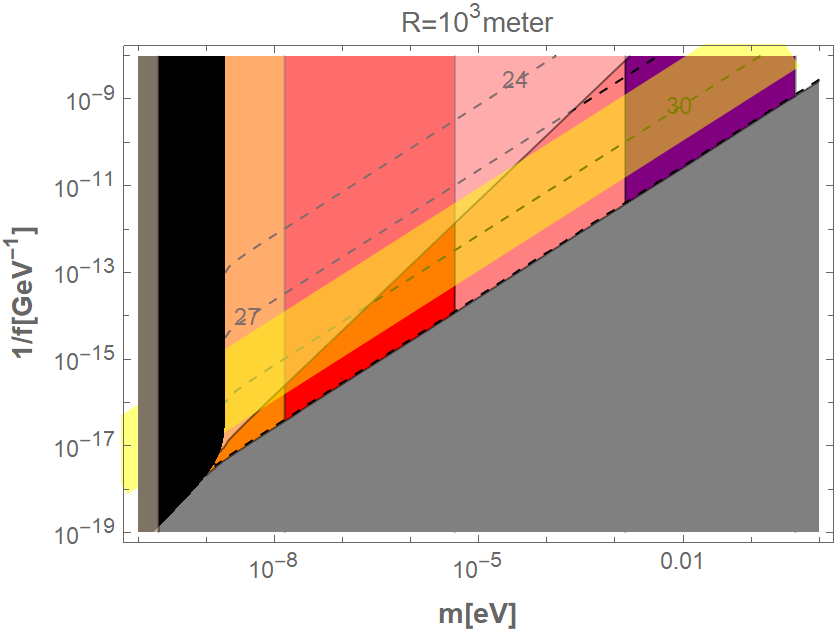}
\caption{ 
Contour plot of spin-down emission power ($P = 10^{37}\,\mathrm{erg/s}$) 
        for rotating axion-like particle (ALP) stars, computed using a 6th-order expansion 
        of the potential. Each panel displays solutions at fixed stellar radius, with vortex 
        winding number $n=1$. The semi-transparent colored regions follow the same 
        parameter conventions as Fig.~\ref{FigBSRep}. We focus exclusively on the dense 
        axion star branch, as the dilute branch solutions predominantly reside in the 
        super-spinning regime ($a>1$) relevant for PeVatron production. The yellow region 
        specifically denotes QCD axion solutions. The black region denotes the non-existence of the real solution, where the  quantity in the root of Eq. \ref{eqAS} becomes negative.  }
\label{FigAS}
\end{figure*}

The perturbative expansion becomes inadequate when $b/f \sim 1$, requiring consideration of the full non-linear potential. In this regime, the system admits an additional solution branch corresponding to dense axion stars with higher central densities. The complete numerical analysis of dense axion stars remains computationally challenging, and their stability properties continue to be debated in the literature \cite{Braaten:2015eeu, Visinelli:2017ooc}. As shown in Fig.~1 of \cite{Chavanis:2017loo}, these solutions exhibit alternating stability patterns at increasing densities. 

Following \cite{Chavanis:2017loo}, we restrict our analysis to the first stable branch by truncating the potential expansion at sixth order. Analogous to the BS case, we assume the critical mass scales linearly with winding number $n$, leading to the generalized mass-radius relation for axion stars:
\begin{equation}
    \tilde{M} = (1+n)^2 \frac{\tilde{R}^3(1+\tilde{R}^2) + \sqrt{\tilde{R}^6(1+\tilde{R}^2)^2 - 48\delta \tilde{R}^4}}{12\delta},
    \label{eqAS}
\end{equation}
where the dimensionless variables are defined as:
\begin{equation}
    \tilde{M} = \frac{Mm}{11M_{\rm pl}f}\,, \,\,\,
    \tilde{R} = \frac{Rmf}{0.2M_{\rm pl}}\,\,\, , 
    \delta = \frac{40f^2}{M_{\rm pl}^2}.
\end{equation}
Using this relation with Eq.~\eqref{eqP}, we calculate the emission power spectrum shown in Fig.~\ref{FigAS}, treating the axion mass $m$ and decay constant $f$ as independent parameters. Our analysis reveals:

\begin{itemize}
    \item Significant parameter space exists for PeVatron emission from gauged axion stars with DM millicharges $q \lesssim 10^{-20}$ and mass $m \sim 10^{-8}$--$1$eV.
    \item Viable sources include axion stars with masses  $M \sim 10^{25}$--$10^{35}$g. 
    \item Dense axion stars satisfying the Kerr bound ($M \sim 10^{30}$ g, $R \sim 10^3$ m) can produce PeVatron emission for $m \gtrsim 10^{-7}$ eV.
\end{itemize}

This behavior differs markedly from that of dilute axion stars (which are governed solely by $b^4$ interactions, as shown in Fig.~\ref{FigBSRep}), demonstrating how higher-order terms in the potential expansion significantly modify the accessible parameter space for PeVatron production. These parameter ranges suggest challenging but potentially detectable signatures for next-generation experiments.

\subsection{Gauged Q-ball}

\begin{figure*}[htbp]
\includegraphics[scale=0.7]{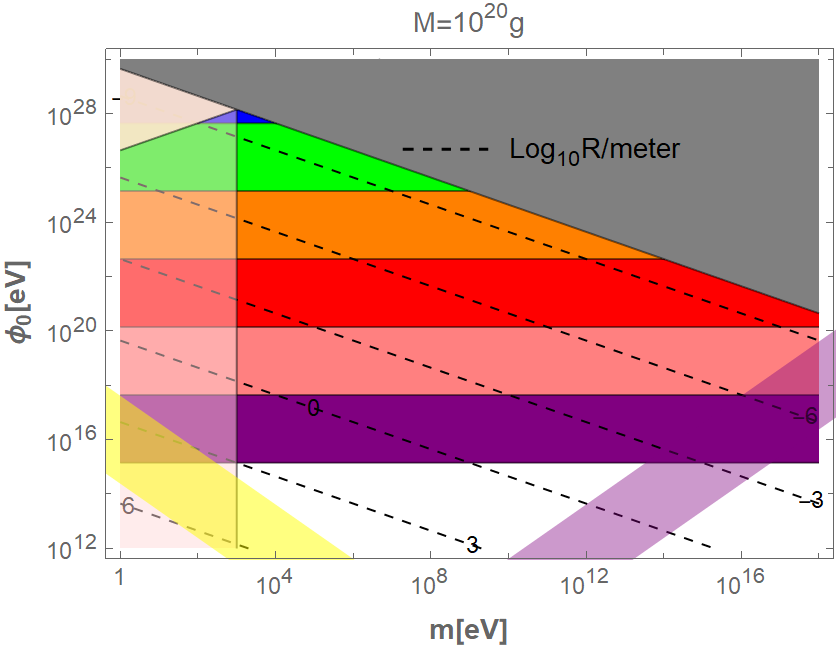}
\includegraphics[scale=0.7]{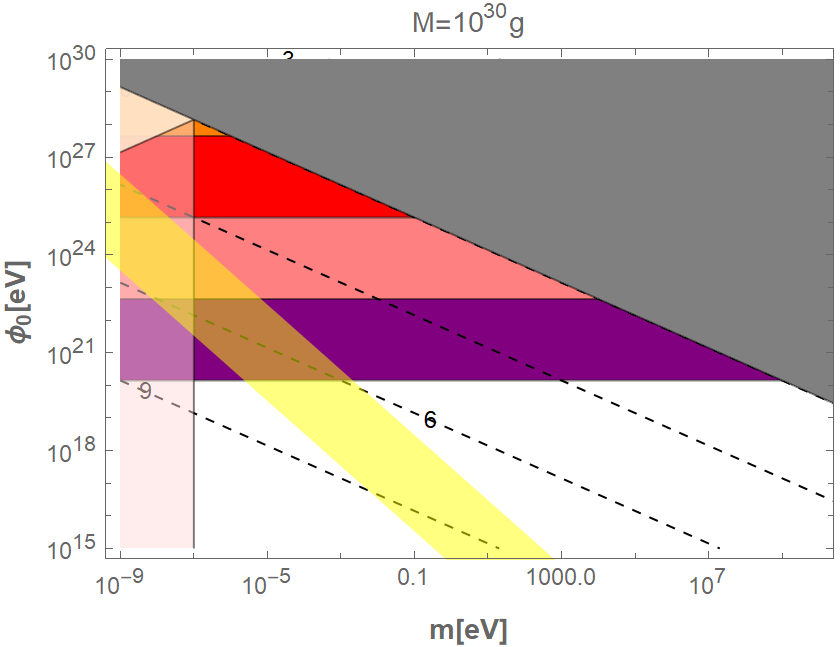} 
\caption{ 
        Contour plot of the spin-down emission power $P = 10^{37}\,\mathrm{erg/s}$ for rotating Q-balls with vortex winding number fixed at $n=1$. The transparent colored regions maintain the same parameter correspondence as in Fig.~\ref{FigBSRep}. The transparent purple shade identifies Higgs-like Q-ball solutions with characteristic field amplitude $\phi_0 \sim m$, while the yellow region represents QCD axion Q-balls modeled using a sixth-order potential expansion. This comprehensive representation captures the essential features of Q-ball emission across different theoretical scenarios while maintaining consistent comparison with previous BS results.} 
\label{FigQball}
\end{figure*}

In flat spacetime where gravitational effects are negligible, stable soliton solutions known as Q-balls can form when bosonic self-interactions precisely balance quantum kinetic pressure. As established in \cite{Coleman:1985ki}, a simple $\phi^4$ potential cannot support such localized configurations. The minimal potential allowing Q-ball formation takes the non-renormalizable form \cite{Friedberg:1986tq}:
\begin{equation}
    V(\phi) = m^2|\phi|^2\left(1 - \frac{|\phi|^2}{\phi_0^2}\right)^2,
    \label{eq-V-Qball}
\end{equation}
which serves as an effective field theory description. This potential features degenerate vacua at $\phi=0$ and $\phi=\phi_0$, with the intervening energy barrier providing the surface tension necessary to stabilize the soliton. The resulting Q-ball configuration has $\phi=\phi_0$ in its interior and $\phi=0$ asymptotically.
In our case, such a scalar field correspond to a Higgs field for abelian gauge $U(1)$,
with the same Lagrangian of Eq.\ref{Abel}.

For rotating Q-balls, we generalize the mass-radius relation from the static case to:
\begin{equation}
    M = \frac{mR^2\phi_0^2}{1+n},
    \label{eq-Qball}
\end{equation}
where $n$ is the vortex winding number. Following our previous approach, we maintain the linear scaling $M_{\rm crit} \propto (1+n)$ in the Schwarzschild limit. Combining this with Eq.~\eqref{eqP} yields the emission power shown in Fig.~\ref{FigQball}, mapped across the $(m,\phi_0)$ parameter space.

Key findings include:
\begin{itemize}
    \item PeVatron production occurs for Q-balls with $M \lesssim 10^{20}$ g and $R \lesssim 10^{-6}$ m.
    \item The millicharge requirement is relatively mild $q \lesssim 10^{-15}$, which however cannot be too large due to stability conditions. In particular, the green area denoting the charge of millicharged DM that falls within current experimental sensitivities \cite{Caputo:2021eaa}.
    \item Higgs-type Q-balls ($\phi_0 \sim m$) favor heavy DM masses $m \sim 10^{15}$--$10^{20}$ eV. Thus, the Yukawa radius of the electromagnetic interaction is, inside the Q-ball, much shorter than the object size. 
\end{itemize}

Notably, the structural similarity between Eqs.~\eqref{eq-V-AS} and \eqref{eq-V-Qball} permits construction of axionic Q-balls using the 6$^{th}$-order potential expansion. However, these non-gravitational counterparts to axion stars can only achieve PeVatron emission when rotating beyond the Kerr bound, representing an extreme regime of the parameter space
which can lead instabilties.

\subsection{Winding number}

\begin{figure*}[htbp]
\includegraphics[scale=0.47]{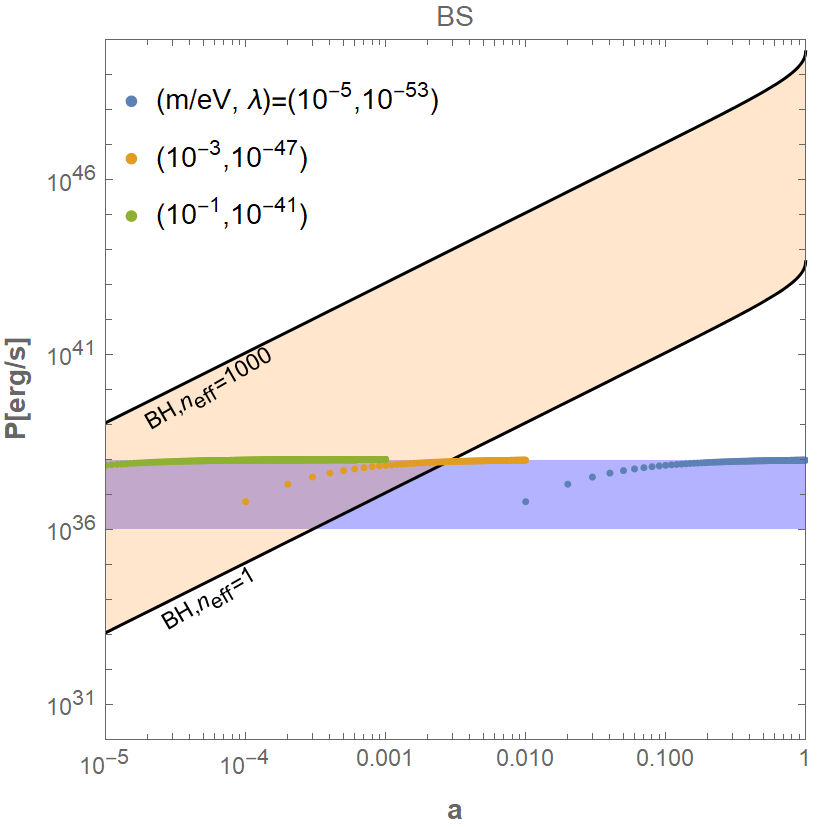}\includegraphics[scale=0.47]{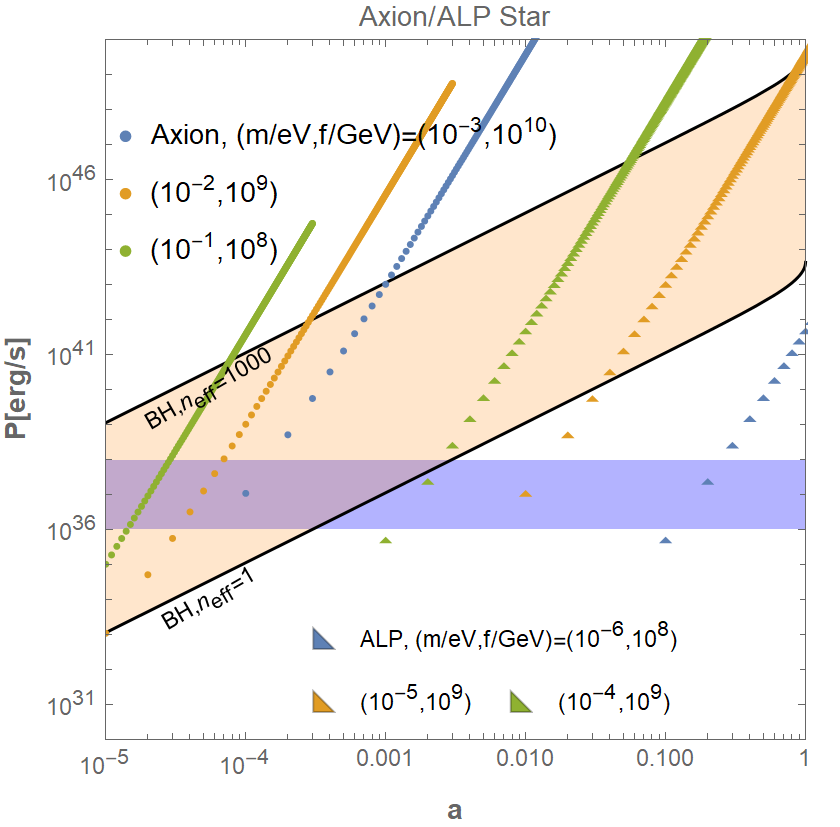}\includegraphics[scale=0.47]{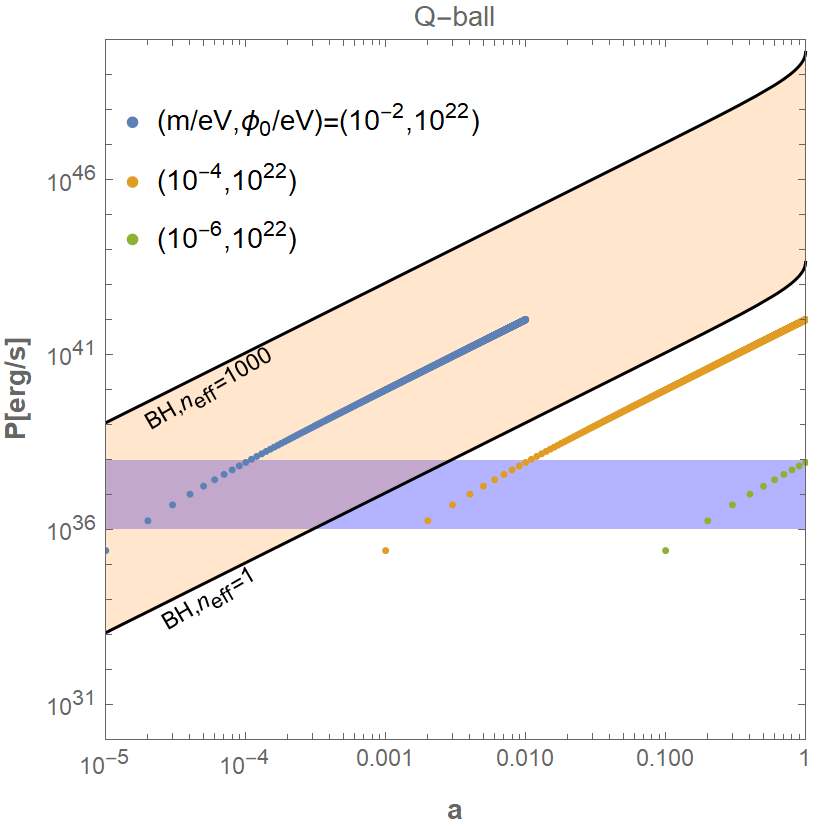}
\caption{ 
Emission power $P$ as a function of dimensionless spin parameter $a$ for rotating BSs, ALP stars and Q-balls  composed of millicharged dark matter. All configurations share common parameters $M = 10^{30}$ g and $q = 10^{-35}$. The orange band represents black hole vortices coupled to millicharged DM, exhibiting a continuous spectrum characterized by the effective flux parameter $n_{\rm eff}$. The purple band indicates the observational window for detected PeVatron emission at $P \approx 10^{37}$ erg/s. This comparative analysis highlights the distinct spin-dependent emission characteristics across different exotic compact object scenarios.}
\label{FigPvsa}
\end{figure*}

In previous sections, we have explored PeVatron production by various ultra-compact soliton objects, restricting our analysis to winding number $n=1$. Here, we extend this investigation to higher winding numbers and examine their impact on spin-down emission power. 

Figure~\ref{FigPvsa} presents the emission power $P$ as a function of the reduced angular momentum $a = J/(GM^2)$ for BSs, gauged axion stars, and gauged Q-balls. A key observation across all soliton cases is the discrete spectrum corresponding to different winding numbers $n$. 

We find the following distinctive scaling behaviors:
\begin{itemize}
    \item BS with repulsive interactions (Higgsed-phase) show emission power largely independent of $a$. 
    \item Gauged axion stars display rapidly increasing emission power scaling as $P \propto a^{6}$.
    \item Gauged Q-balls follow a quadratic scaling $P \propto a^{2}$.
\end{itemize}

For comparison, we also include the case of a black hole vortex coupled to a millicharged dark matter cloud. While experimentally challenging, the discrete PeVatron spectrum and the characteristic $P(a)$ scaling relations could potentially serve as diagnostic tools to differentiate between various exotic compact object (ECO) sources.

\subsection{Gravitational radiation}

\begin{figure*}[htbp]
\includegraphics[scale=0.46]{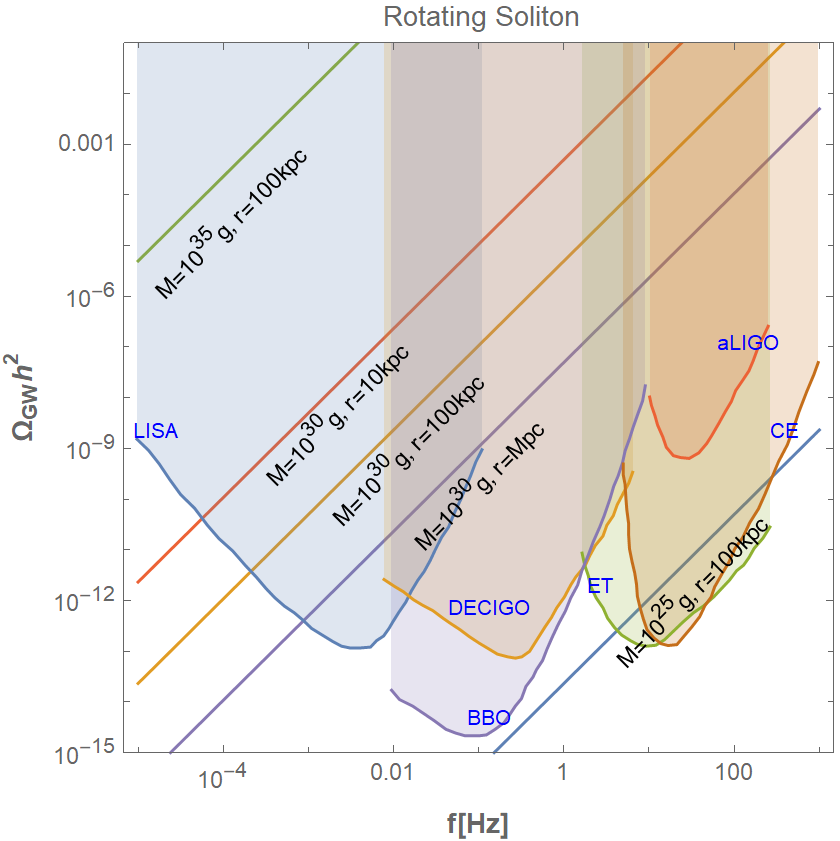}
\includegraphics[scale=0.46]{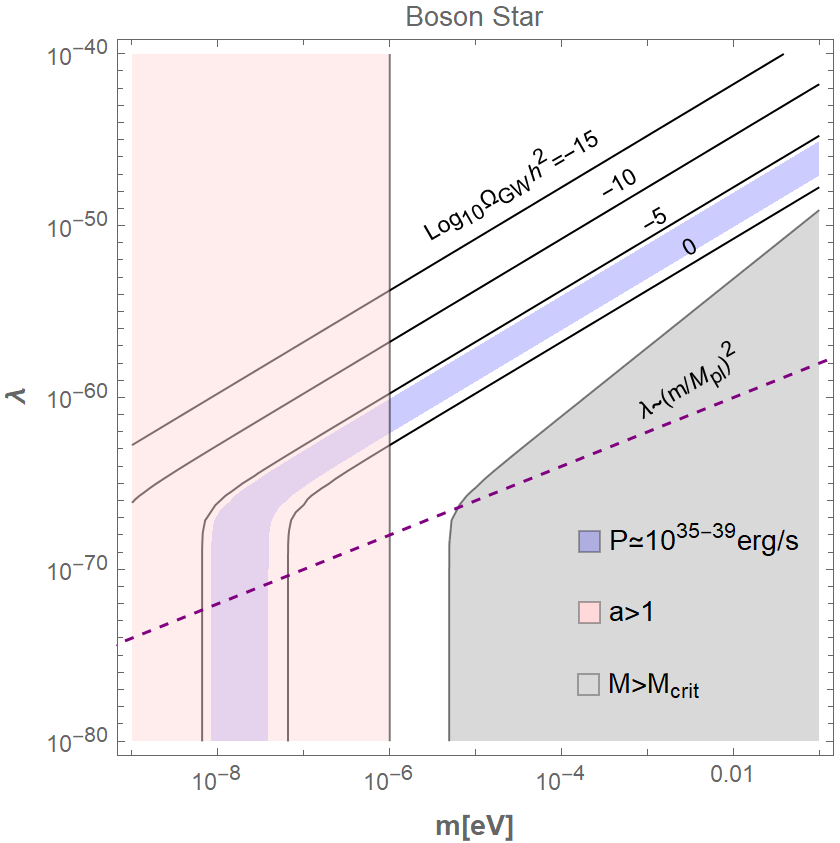} 
\includegraphics[scale=0.46]{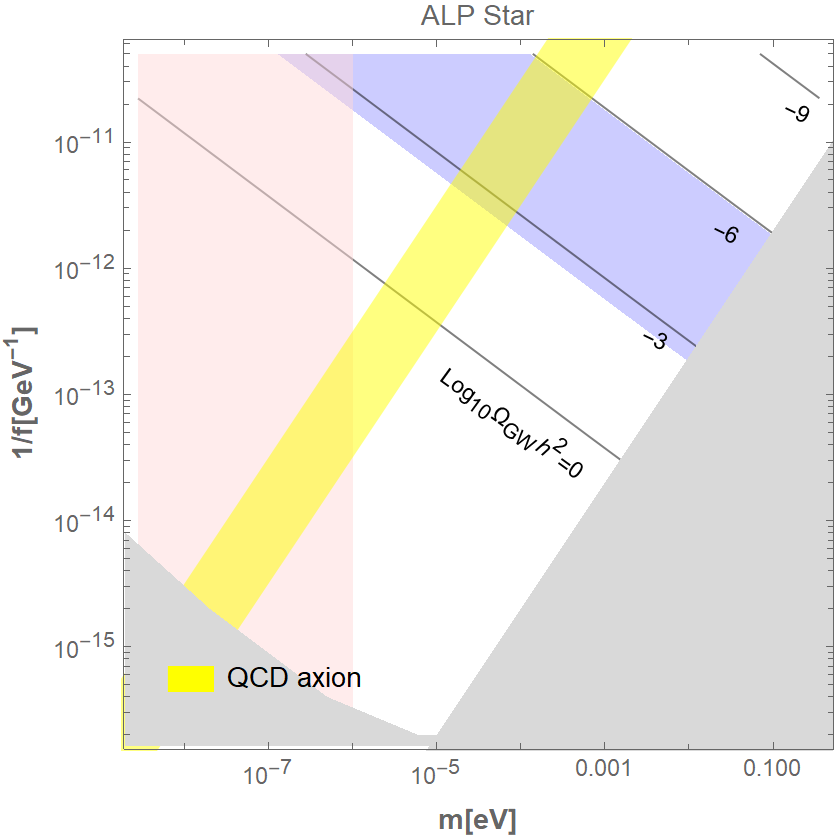}
\caption{ 
        \textbf{Left:} Gravitational wave emission spectrum from individual rotating soliton objects, showing the dependence on mass $M$ and distance $r$ from Earth. \textbf{Other panels:} Contour plots of gravitational wave abundance $\Omega_{\rm GW}$ in the parameter space of rotating BSs and ALP stars. The light blue shaded regions indicate millicharged dark matter emission in the PeVatron energy range $P \simeq 10^{35}$--$10^{39}$ erg/s. All calculations assume fixed parameters: object mass $M = 10^{30}$ g, distance $r = 100$ kpc, vortex winding number $n = 10$, and millicharge parameter $q = 10^{-30}$.}
\label{FigGW}
\end{figure*}

As demonstrated in \cite{Ferrell:1989kz}, rapidly rotating BSs in high-spin configurations represent excited states with non-zero quadrupole moments, enabling gravitational wave (GW) emission through transitions to lower-energy states. The fundamental process involves quantized changes in the vortex winding number ($\Delta n = 2$) with graviton emissions. 

The toroidal topology of constant energy density surfaces in rotating BS \cite{Mielke:2016war,Siemonsen:2023hko} suggests an approximate treatment as an inspiraling binary system. We model a rotating BS of mass $M$ and radius $R$ as two equal-mass ($M/2$) components separated by $2R$, yielding the GW energy density as
\begin{equation}
    \Omega_{\rm GW} = 0.1\left(\frac{M}{10^{30}\,\mathrm{g}}\right)^{5/3}\left(\frac{\mathrm{kpc}}{r}\right)^2\left(\frac{f_{\rm GW}}{\mathrm{Hz}}\right)^{5/3},
    \label{eq-BSGW}
\end{equation}
where $r \sim \mathcal{O}(100\,\mathrm{kpc})$ represents typical Milky Way halo scales, and $f_{\rm GW} = \Omega/\pi$ connects to the BS rotation.

Figure~\ref{FigGW} displays the GW spectrum for various BS masses and distances, compared with sensitivity curves of upcoming detectors (aLIGO, CE, ET). Our fiducial analysis assumes soliton mass $M = 10^{30}$ g at $r = 100$ kpc\, DM Millicharge $q = 10^{-30}$. Specifically, we consider two Two typical cases with repulsive $\phi^4$ BS and dense Axion stars with $\phi^6$ potential truncation. The key findings include:
\begin{itemize}
    \item Concurrent PeVatron ($P \sim 10^{35-39}$ erg/s) and GW emission,($\Omega_{\rm GW}h^2 \sim 10^{-5}-10^{-3}$).
    \item Characteristic frequencies $f_{\rm GW} \sim 1-100$ Hz fall within the aLIGO/CE/ET bands.
\end{itemize}

This multi-messenger signature - combining PeV gamma rays and intermediate-frequency GWs - provides a unique probe of ECOs beyond standard astrophysical objects.

\section{Discussion and Conclusions }

In this study, we have explored the theoretical framework and observational implications of \textit{exotic PeVatrons}---novel astrophysical sources capable of producing UHE gamma rays beyond the PeV scale. These ECOs, including ultra-spinning black holes vortex, BSs, Q-balls, and gauged ALP stars, are hypothesized to arise from quantum gravity and light millicharged DM scenarios. Their unique properties, such as extreme rotational dynamics, millicharged DM interactions, and quantized magnetic flux, enable particle acceleration mechanisms that surpass the limits of conventional astrophysical sources like pulsar wind nebulae and supernova remnants.

Our analysis demonstrates that such ECOs can generate powerful PeVatronic emissions through mechanisms like vortex-string systems, superradiant instabilities, and nonlinear self-interactions. For instance, rotating BSs and black hole vortices coupled to millicharged DM can produce UHE gamma rays with luminosities approaching $10^{37}\,\text{erg/s}$, detectable by current such as  LHAASO \cite{LHAASO:2019qtb}, HAWC \cite{Tibolla:2023sdj}) and future (e.g., CTA \cite{CTAConsortium:2023tdz}) observatories. 
The distinct spectral signatures and discrete emission power scaling with winding number $n$ provide potential observational discriminant between different ECO classes.

Moreover, these objects may exhibit multi-messenger signatures, including GWs from high-spin configurations, offering complementary probes into their nature. The predicted GW frequencies ($1$--$100\,\text{Hz}$) fall within the sensitivity ranges of upcoming detectors like aLIGO, CE, and ET.

While challenges remain---such as uncertainties concerning the distribution of these objects in the dark halo---our work underscores the potential of Exotic PeVatrons to bridge particle physics, astrophysics, and cosmology. Future observations of UHE gamma rays beyond $10\,\text{PeV}$ could provide unambiguous evidence for these objects, opening a new window into physics beyond the Standard Model and the nature of dark matter.

The Exotic PeVatron paradigm can also be related to new mechanism for UHE Neutrinos as observed by KM3NeT collaboration \cite{KM3NeT:2025npi}. Indeed, contrary to photons, neutrinos can open the window of extra-galactic Exotic PeVatrons. 
Moreover, the AUGER collaboration has established upper limits on ultra-high-energy (UHE) gamma rays and neutrinos with energies beyond approximately $50 -- 100{\rm PeV}$. These results rule out certain exotic Pevatron models, though only within specific, model-dependent regions of their parameter space \cite{Gonzalez:2025bgx}.

In summary, exotic PeVatrons represent a compelling frontier in high-energy astrophysics, with profound implications for understanding the Universe's most energetic phenomena and the fundamental laws governing them.

\vspace{0.1cm}

{\bf Acknowledgements}
The Authors  thank Luca Visinelli for useful discussions and suggestions on the topics.
The work of A.A.\ is supported by the National Science Foundation of China (NSFC) 
through the grant No.\ 12350410358; 
the Talent Scientific Research Program of College of Physics, Sichuan University, Grant No.\ 1082204112427;
the Fostering Program in Disciplines Possessing Novel Features for Natural Science of Sichuan University, Grant No.2020SCUNL209 and 1000 Talent program of Sichuan province 2021.
S.C.  and QG acknowledge the Istituto Nazionale di Fisica Nucleare (INFN) Sez. di Napoli,  Iniziative Specifiche QGSKY and MoonLight-2. SC acknowledges  the Istituto Nazionale di Alta Matematica (INdAM), gruppo GNFM, for the support.
This paper is based upon work from COST Action CA21136 -- Addressing observational tensions in cosmology with systematics and fundamental physics (CosmoVerse), supported by COST (European Cooperation in Science and Technology).

\end{document}